\documentclass[twocolumn,prl,showpacs,amsmath,amssymb,superscriptaddress,hyperref]{revtex4}

\usepackage{graphicx}
\usepackage{dcolumn}
\usepackage{bm}

\begin{document}


\title{Enhancing single-molecule photostability by optical feedback\\
 from quantum-jump detection}

\author{Vincent Jacques} 
\author{John Murray}
\affiliation{Laboratoire de Photonique Quantique et Mol\'eculaire ENS Cachan , UMR CNRS
8537, 61 avenue du Pr\'esident Wilson, 94235 Cachan cedex, France}
\author{Fran\c{c}ois Marquier}
\affiliation{Laboratoire EM2C, Ecole Centrale Paris, Grande Voie des Vignes, 92295 Chatenay-Malabry cedex, France}
\author{Dominique Chauvat}
\author{Fr\'ed\'eric Grosshans}
\author{Fran\c{c}ois Treussart}
\author{Jean-Fran\c{c}ois Roch}
\email{roch@physique.ens-cachan.fr}
\affiliation{Laboratoire de Photonique Quantique et Mol\'eculaire ENS Cachan , UMR CNRS
8537, 61 avenue du Pr\'esident Wilson, 94235 Cachan cedex, France}

\date{\today}

\begin{abstract}
We report an optical technique that yields an enhancement of single-molecule photostability, by greatly suppressing photobleaching pathways which involve photoexcitation from the triplet state. This is accomplished by dynamically switching off the excitation laser when a quantum-jump of the molecule to the triplet state is optically detected. This procedure leads to a lengthened single-molecule observation time and an increased total number of detected photons. The resulting improvement in photostability unambiguously confirms the importance of photoexcitation from the triplet state in photobleaching dynamics, and may allow the investigation of new phenomena at the single-molecule level.
\end{abstract}



\maketitle

\indent During the last ten years, optically-based single-molecule detection has become a widely used technique ~\cite{review,Xie,Moerner_review} that has revealed many phenomena hidden in ensemble-averaged experiments, ranging from single-emitter effects in quantum optics~\cite{Orrit_tenyears,Lounis,Sandoghdar} to insights in biophysics~\cite{Yanagida,Weiss_science,XieBis,GFP}. However, photobleaching, i.e., the irreversible conversion of an optically excited organic fluorophore into a non-fluorescent entity, is a severely limiting factor in all single-molecule studies realized under ambient conditions. There is strong experimental evidence that the main photobleaching pathways begin with the molecule being in the triplet state $T_{1}$~\cite{Orrit,Widengren}, which is a metastable dark state to which the molecule has a small probability to jump through intersystem crossing (ISC)~\cite{Lakowicz}. Indeed, photostability can be greatly improved by engineering organic fluorophores with intrinsically low ISC rates~\cite{Margineanu,Mullen}. Moreover, some of these pathways involve further photoexcitation from this triplet state~\cite{Eggeling,Deschenes}. Based on these results, an obvious way to decrease photobleaching is to avoid any light excitation while the molecule is in the triplet state. 
\indent For an ensemble of molecules, this can be achieved by a pulsed excitation, with a pulse duration shorter than the excited-state lifetime and a repetition period longer than the triplet lifetime $\tau_{\rm t}$. For each excitation pulse, molecules will undergo a single fluorescence cycle and if ISC occurs, they will decay back to the ground state $S_{0}$ before the following excitation pulse. This strategy is routinely used in high-power dye lasers~\cite{dye_laser} and has recently been applied in the context of high-resolution fluorescence microscopy~\cite{Hell}. However, in practice, this method is not compatible with single-molecule detection since the maximum excitation rate of the order of $1/\tau_{\rm t}$ corresponds to a photon flux much below the sensitivity threshold allowing single-molecule detection. 
\indent We report a new technique to strongly reduce photobleaching that has intrinsic single-molecule sensitivity, and which works without the synchronization of the molecule to a pulsed excitation. The scheme consists of a feedback loop based on the real-time detection of quantum-jumps from the singlet state $S_{1}$ to the triplet state $T_{1}$. \\

\begin{figure}[t]  
\centerline{\includegraphics[width=9cm]{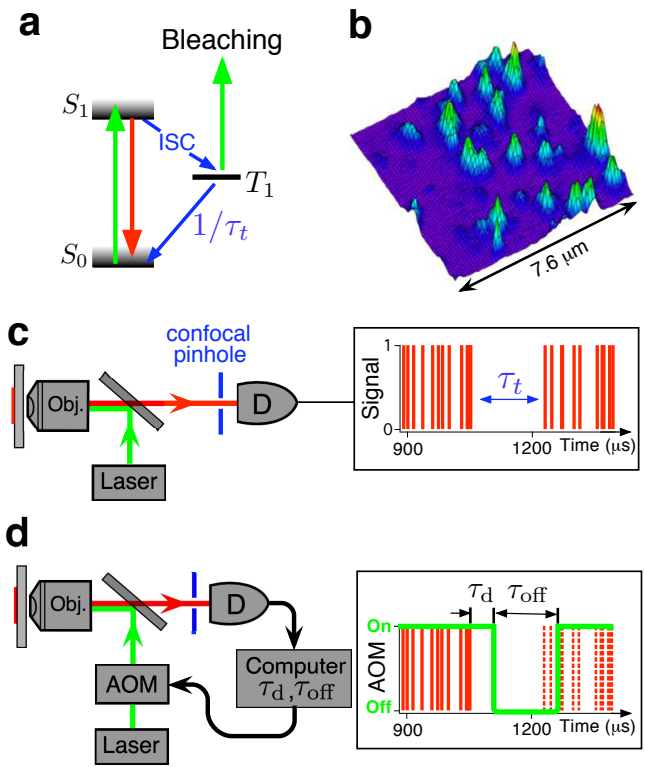}}
\caption{Principle of the experiment. \textbf{(a)} Single-molecule energy levels in the three-level Perrin-Jablonski diagram. Photons emitted through the $S_{1}\rightarrow S_{0}$ transition are detected with a standard confocal microscope \textbf{(c)}, as shown by a typical raster scan of the sample \textbf{(b)}. The single-molecule fluorescence time trace \textbf{(c)} reveals dark periods with durations on the order of the triplet-state lifetime $\tau_{\rm t}$, corresponding to quantum-jumps to the triplet state $T_1$ by inter-system crossing (ISC). The displayed single-molecule fluorescence signal represents the output of the photon-counting detector (D) without any binning. {\bf (d)} The quantum-jump detection triggers an acousto-optical modulator (AOM), used as an optical switch for the excitation laser. If no photon is detected during a time window of duration $\tau_{\rm d}$, the laser is switched off for a duration $\tau_{\rm off}$ longer than $\tau_{\rm t}$. The graph shows in green the AOM command for the fluorescence time trace represented in \textbf{(c)}. The characteristic response time of the AOM is $600 \ \rm ns$, much shorter than all other time constants of the experiment.}
 \end{figure}
 
\indent Figure 1 shows the principle of the experiment using a simple model for fluorescence. Under light irradiation, light-emission from an organic fluorophore  can be described using the usual three-level Perrin-Jablonski representation~\cite{Lakowicz}, with cycles occurring between the singlet ground state $S_{0}$ of the molecule and its first excited singlet state $S_{1}$  (Fig.1-(a)). Even though the intersystem crossing (ISC) from the $S_{1}$ state to the first triplet state $T_{1}$ is spin-forbidden, spin-orbit interaction leads to a non-radiative decay of the molecule to this level with a small probability. As relaxation from $T_{1}$ to $S_{0}$ is also spin-forbidden, the triplet-state lifetime $\tau_{\rm t}$ is orders of magnitude larger than the one associated with fluorescence.
The experimental setup is based on a home-made confocal microscope optimized for single-molecule fluorescence detection under ambient conditions. Samples are made by spin-coating on a glass coverslip a solution of polymethyl methacrylate (PMMA, $1\%$ mass in anisole) doped with dye molecules at nanomolar concentrations. The result is a few tens of nanometers thick polymer coating, in which molecules are well separated and can be individually probed~\cite{alleaume}.

\indent Fluorescence from a single molecule is detected by an avalanche photodiode operated in the photon-counting regime (Fig.1-(b,c)). Due to the long lifetime of the triplet state, a non-radiative decay from $S_{1}$ to $T_{1}$ appears as a sudden drop in the fluorescence signal, corresponding to a quantum-jump in the single-molecule emission time trace~\cite{Basche_nature,VanHulst} (Fig. 1-(c)). The fluorescence signal is fed into an electronic card that commands in real-time an acousto-optical modulator to switch off the excitation laser when a quantum-jump is detected (Fig. 1-(d)). The result is a quantum-jump-based feedback loop that performs an adaptation of light excitation to the single-molecule dynamics with two adjustable time constants, $\tau_{\rm d}$ and $\tau_{\rm off}$. As described in Fig.1-(d), $\tau_{\rm d}$ is the time constant over which the decision is taken that a molecule has indeed experienced a $S_1\rightarrow T_1$ quantum-jump. This parameter acts as a temporal threshold that discriminates between fluorescence cycles and a quantum-jump: If no photon is detected during $\tau_{\rm d}$, the laser is switched off for a duration $\tau_{\rm off}$ which is set to be greater than the triplet lifetime $\tau_{\rm t}$.\\

\begin{figure} [b]  
\centerline{\includegraphics[width=7cm]{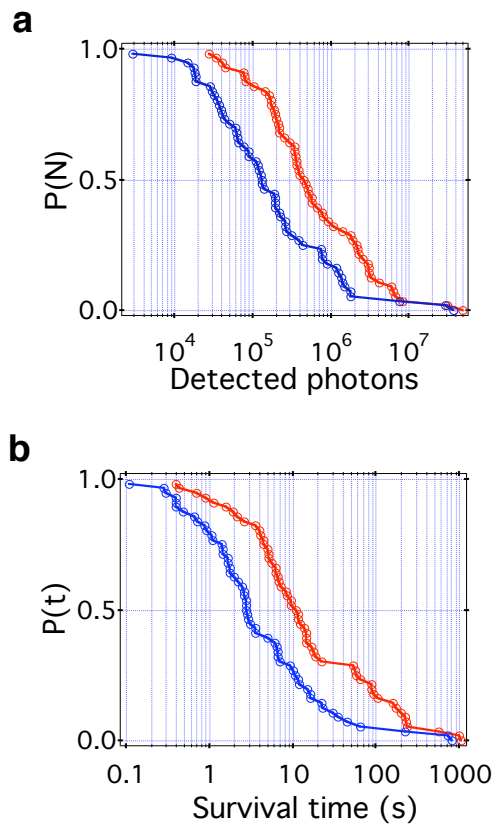}}
\caption{ Photostability enhancement: Experimental results for DiI represented as accumulated probability distributions. $P(N)$ (resp. $P(t)$) is the probability that a single molecule has not photobleached before the detection of $N$ photons emitted by the molecule (resp. before a survival time $t$). The statistics correspond to a set of 56 molecules without the quantum-jump feedback loop (blue points) and another set of 56 molecules with the feedback loop running (red points) with $\tau_{\rm d}=70 \ \mu \rm s$ and $\tau_{\rm off}=400 \ \mu \rm s$. The inferred probability distributions show an enhancement of both the total number of recorded photons (\textbf{a}) and the survival time (\textbf{b}) before photobleaching, when the feedback loop is applied. }
\end{figure}

\indent The experiment is first performed with the carbocyanine DiI, an ionic dye commonly used in biology as a label for probing electric potential and lipid lateral mobility in membranes~\cite{Axelrod}. Single molecules are excited with a cw laser at 532~nm wavelength with an intensity of $1  \  \rm kW.cm ^{-2}$, close to the saturation of the $S_{0} \rightarrow S_{1}$ transition. This leads to detection counting rates ranging from 30 to 200~kcounts.s$^{-1}$, depending on the environment and the molecule dipole orientation. To limit the unavoidable bias which appears in single-molecule statistics measurements, we decide to study all fluorophores with counting rates higher than 30 kcounts.s$^{-1}$, equivalent to a mean time interval between two consecutive detection events of $\tau_{\rm fluo}= 33 \  \mu \rm s$. In order to prevent permanent switching of the excitation laser, the time constant $\tau_{\rm d}$ must be greater than $\tau_{\rm fluo}$ and is set at $70 \ \mu \rm s$. Moreover, the triplet-state lifetime $\tau_{t}$ is approximately $200 \ \mu \rm s$ for the DiI molecule. As $\tau_{t}$ varies depending on the local environment~\cite{VanHulst}, the other time constant of the feedback loop $\tau_{\rm off}$ is set at $400 \ \mu \rm s$.\\
\indent For an ensemble of single molecules, we measure the total number of detected photons before photobleaching with and without the quantum-jump-based feedback loop. This parameter is directly related to the amount of information that can be extracted from a single molecule. We then evaluate from this set of data the probability $P\left(N\right)$ that a molecule has not photobleached before the record of $N$ photocounts. The resulting probability distributions, plotted in Fig. 2-(a), demonstrate a photostability enhancement when the feedback loop is applied. A Kolmogorov-Smirnov test~\cite{NumRec} is then performed in order to verify that the difference observed between the data sets is not due to statistical fluctuations but indeed to different underlying probability distributions. According to this test, the probability to observe such different statistics  from an identical probability distribution is as small as $2.2\times 10^{-4}$. This confirms the effectiveness of the feedback loop to a high degree of confidence. In addition, we measure the survival time before photobleaching~\cite{NoteSurvivalTime}, which is also enhanced by applying the feedback loop (Fig. 2-(b)). We note the existence of a subset of molecules with long-lasting fluorescence, that remain unaffected by the feedback loop. This observation is consistent with the results of Ref.\cite{VanHulst} which reports the distribution of DiI triplet-state lifetimes in a thin PMMA layer. In this reference, $\tau_{\rm t}$ is smaller than $70\  \mu \rm s$ for approximatively $10\%$ of the molecules. For that subset of molecules, which are the most photostable ones, the quantum-jump-based feedback loop with $\tau_{\rm d}=70 \ \mu \rm s$ will obviously be ineffective, as observed.\\

\begin{figure} [t]  
\centerline{\includegraphics[width=6cm]{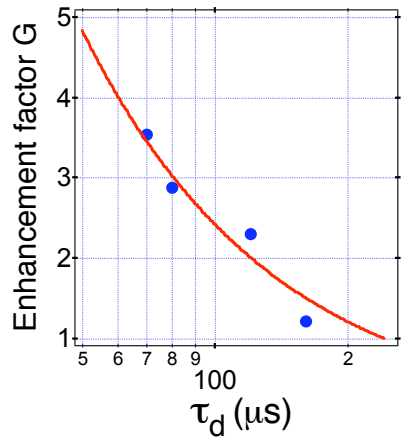}}
\caption{ Photostability enhancement factor $G$ as a function of the time constant $\tau_{\rm d}$ of the feedback loop.  For each value of $\tau_{\rm d}$, the statistics of the numbers of detected photons are measured for 50 molecules, with and without the feedback loop. The value of $G$ factor is evaluated by dividing the medians of the corresponding sets of data. The maximum value $G=3.6$ is obtained for the probability distributions displayed in Fig. 2-(a). The solid line represents the value of $G$ factor given by the model  with $<\tau_{\rm t}>=240 \ \mu \rm s$. }
 \end{figure}
 
\begin{figure} [b]  
\centerline{\includegraphics[width=7cm]{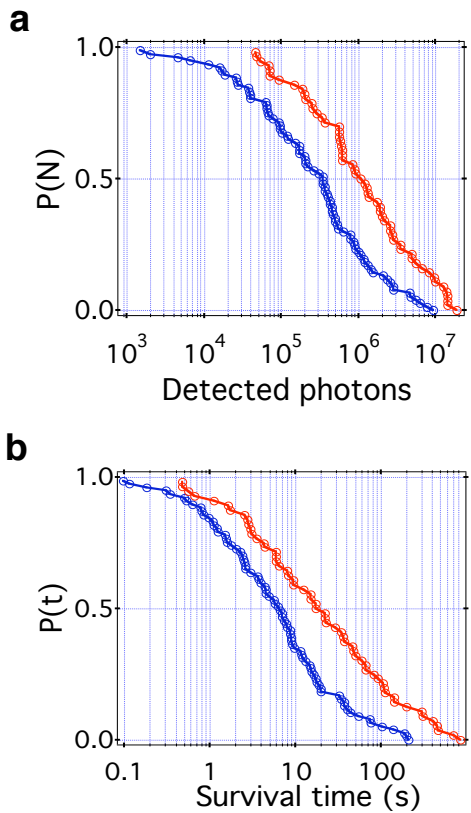}}
\caption{ Photostability enhancement for terrylene, corresponding to $G=3.1$. Statistics are measured with a set of 72 molecules without the quantum-jump-based feedback loop (blue points) and another set of 57 molecules running the feedback loop (red points). Time $\tau_{\rm d}$ of the feedback loop is set at $70 \ \mu \rm s$ and time $\tau_{\rm off}$ at $400 \ \mu \rm s$.}
\end{figure}
 
 \indent It is also desirable to quantify the gain in photostability. We tentatively introduce a simple model in which all photobleaching pathways require further photoexcitation from the triplet state. In this model, there is no photobleaching when the triplet state is empty or when the excitation laser is switched off. After a quantum-jump, the probability for the molecule to photobleach is proportional to the duration for which it remains under laser irradiation in the triplet state. The effect of the quantum-jump-based feedback-loop is a reduction of this illumination duration from $\tau_{\rm t}$ to $\tau_{\rm d}$ (Fig. 1-(d)). We then expect a photostability enhancement $G$ on the order of 
 
 \begin{equation}
 G= \frac{<\tau_{\rm t}>}{\tau_{\rm d}}
\end{equation}
where $<\tau_{\rm t}>$ is the average value of the triplet-state lifetime. To test this dependance, the experiment is performed for different values of $\tau_{\rm d}$. For each ensemble of molecules, triplet-state lifetimes are distributed with complex statistics~\cite{VanHulst}, and will then correspond to different photostability gains. Therefore, for each values of $\tau_{\rm d}$ we choose to estimate $G$ as the ratio of the medians of the photocounts distributions measured with and without the feedback loop. The median is known as a robust estimator of the central value of the distribution of a random variable~\cite{NumRec}, even for heavy-tailed statistics that have been found in photobleaching ensemble-measurements~\cite{Didier}. The experimental results show an increase of the $G$ factor as $\tau_{\rm d}$ is decreased (Fig. 3). The simple model matches the data well for  $<\tau_{\rm t}>=240  \ \mu \rm s$, consistent with previous measurements~\cite{VanHulst,alleaume}. This clearly demonstrates the important role of photoexcitation from the triplet state in photobleaching pathways.\\
 \indent As is well known, an important factor in photobleaching dynamics is the ambient atmosphere, especially the presence of oxygen~\cite{Orrit,Lill} which can lead to photobleaching by oxidation of the dye. However, the effect of oxygen changes depending on the molecule. Whereas oxygen acts as a quencher of the triplet state for DiI, leading to an increase of the survival time before photobleaching, it enhances photodestruction in the case of aromatic hydrocarbon dyes such as  terrylene~\cite{Christ,Renn}. To investigate a potential generality of the quantum-jump-based feedback method, the experiment is reproduced with terrylene molecules dispersed in a thin layer of PMMA (Fig. 4). An enhancement of both the total number of emitted photons and the survival time is again achieved, when the feedback loop is applied. The enhancement factor $G$ is estimated following the same procedure as above. We find $G=3.1$, again in good agreement with the ratio $<\tau_{\rm t}>/\tau_{\rm d}$  for terrylene.\\
 
 \indent We have shown that a simple adaptative light excitation scheme, relying on the real-time detection of single-molecule quantum-jumps, allows to greatly enhance the photostability of different types of organic fluorophores that all suffer from photobleaching. This optical technique can be operated at room temperature and under ambient atmosphere. Further photostability improvement is expected by decreasing the threshold parameter $\tau_{\rm d}$ in the feedback loop. This requires an increase in the experimental set-up collection efficiency or a decrease of the fluorescence lifetime, as can be achieved by coupling single-molecule fluorescence to a nanostructured substrate~\cite{Rigneault}.  The achieved photostability enhancement may open new doors in the understanding of single-molecule dynamics.\\

We are indebted to Philippe Grangier for pinpointing the influence of light excitation in the triplet state and to Andr\'e Clouqueur for realizing the electronics of feedback loop. We also acknowledge Jacques Delaire and Robert Pansu for fruitful discussions. This research is supported by Institut Universitaire de France.


\begin{thebibliography}{30}
\bibitem{NoteSurvivalTime}
For each molecule in the set of data, the survival time before photobleaching is corrected for all periods when the laser is switched off.
\end{thebibliography}
 \end{document}